\documentclass[aps,prl,twocolumn,groupedaddress]{revtex4-1}

\usepackage{graphics}
\usepackage{epsfig}
\usepackage{color}

\begin{document}



\title{Beyond Carbon K-edge harmonic emission using spatial and temporal synthesized laser field}

\author{J. A. P\'erez-Hern\'andez$^{1}$}
\email[]{joseap@usal.es}
\author{M. F. Ciappina$^2$}
\author{M. Lewenstein$^{2,3}$}
\author{L. Roso$^{1}$}
\author{A. Za\"ir$^4$}

\affiliation{$^1$Centro de L\'aseres Pulsados CLPU,  E-37008 Salamanca, Spain}
\affiliation{$^2$ICFO-Institut de Ci\`ences Fot\`oniques, Mediterranean Technology Park, 08860 Castelldefels (Barcelona), Spain}
\affiliation{$^3$ICREA-Instituci\'o Catalana de Recerca i Estudis Avan\c{c}ats, Lluis Companys 23, 08010 Barcelona, Spain}
\affiliation{$^4$Department of Physics, Imperial College London, London, SW7 2AZ, United Kingdom}


\date{\today}

\begin{abstract}

We present numerical simulations of high-order harmonic generation in helium using a temporally synthesised and spatially non-homogeneous strong laser field. The combination of  temporal and spatial laser field synthesis results in a dramatic cut-off extension far beyond the usual semi-classical limit. Our predictions are based on the convergence of three complementary approaches:  resolution of the three dimensional Time Dependent Schr\"odinger Equation,  time-frequency analysis of the resulting dipole moment and classical trajectories extraction. Employing a combination of temporally and spatially synthesised laser field provides coherent XUV photons beyond the carbon K-edge which is of high interest for initiating inner-shell dynamics and study time- resolved intra-molecular attosecond spectroscopy. 



\end{abstract}


\maketitle

Near-edge X-ray absorption spectroscopy~\cite{rehr00, bressler04} is a very powerful technique to probe local chemical environment of molecules. Providing to such technique an attosecond time resolution will open a new route to explore inner shell inducing ultra-fast dynamics of charges in molecular system and therefore it is expected to be an important breakthrough. One way to achieve attosecond X-ray is the generation of  high harmonics (HHG) driven by mid-infrared strong laser field, as it is well known that the HHG cut-off scales as $\lambda^2$  ($\lambda$ being the laser wavelength)~\cite{lewenstein94A}.  A first demonstration has been very recently done~\cite{pop12} where high requirement on the mid-infrared wavelength femtosecond laser source may remain quite technologically challenging. In addition, it has been demonstrated that the harmonics efficiency strongly decreases with longer wavelength according to a law in $\lambda^{-5.5}$~\cite{tate_07A, frolov_08, perez-hernandez09A}. Therefore high demands are on developing alternative routes based on existing conventional femtosecond laser sources at $800$ nm.  So far a large amount of strategies to tailor such input laser pulses for HHG have been carried out. Using a combination of two or more colours laser fields to synthesise temporally the final strong laser field~\cite{zeng07,Takahashi08,Winterfeldt08,chipperfield09, chipperfield10, siegel10,brugera10,Orlando2010,brugera11,ganeev12,bandrauk12}; or using chirping techniques~\cite{carrera09}. The new approach we propose consists in  the combination of two techniques to shape the final laser field both in time and in space. The temporal synthesis is obtained using two few-cycle laser pulses delayed in time~\cite{perez-hernandezMD09}. Few-cycle laser pulses are now commonly used and are produced using post-compression stage for attoscience applications~\cite{sansone_s_2006, eckle_np_2008}. The spatial synthesis is obtained by using a non-homogeneous laser field \cite{husakou11,yavuz12, marcelo12} as produced by nano-plasmonic antennas for instance. The goal of this paper is to show that synthesising the laser field both in time and in space is a new route to generate photon via HHG beyond the carbon K absorption edge and this using near infrared wavelength as $800$ nm, which corresponds to conventional femtosecond Ti:Sa laser sources for strong field physics. 

To synthesised temporally the laser field we use a method described in~\cite{perez-hernandezMD09} where two delayed pulses of 4-cycles at 800 nm are combined and correspond to,
\begin{equation}
E_1(t)=E_0 \sin^2\left( { \omega t \over 2N }\right) \sin (\omega t)    
\label{e1}
\end{equation}

\begin{equation}
E_2(t,\tau)=E_0 \sin^2\left( { \omega (t-\tau) \over 2N }\right) \sin (\omega (t-\tau)+\phi)            
\label{e2}
\end{equation}
where $E_0$ is the laser electric field amplitude, in atomic units ($E_0=\sqrt{I/I_0}$ with $I_0=3.5\times 10^{16}$ W/cm$^2$),  $\omega=0.057$ a.u. the laser frequency corresponding to $\lambda=800$ nm, $T$ the laser period, $N$ the total number of cycles in the pulse, $\phi$ the carrier-envelope phase (CEP) and $\tau$ the time delay between the pulses. In our simulations we consider the case $N = 4$ and $\phi=0$.  

According to our numerical simulations the optimal time delay between these two pulses corresponds to $\tau=1.29T$. This way the laser amplitude after the superposition is equal to the two input pulse replicas. The two replica have the same carrier envelop phase (CEP),  which is always the case if the replicas are produced from the same initially CEP stabilised source. With such configuration an extension of the cut-off has been predicted up to $4.5 U_p$, where $U_p$ is the quiver energy associated to the laser field employed ($U_p=\frac{I}{4\omega^2}$). Most of the numerical and semi-classical approaches to study HHG in atoms and molecules consider homogeneous temporal and spatial laser field distributions \cite{protopapas97, brabec00}. Some studies have study the effect of homogeneous temporal distribution together with synthesised spatial profile. In~\cite{strelkov09}  they proposed to synthesised a spatially flat top profile to produce isolated attosecond pulses showing that the spatial domain control of the laser field  provides degree of freedom to control HHG.  

Our idea is therefore to use not only temporally synthesised laser field using a combination of two CEP stabilised few-cycle pulses as described by Eq. (\ref{e1}) and (\ref{e2}) (referred to as double pulse configuration) but also spatially locally synthesised (referred to as non-homogeneous configuration). This non-homogeneous spatial distribution of the laser field can be obtained experimentally by using a non-homogeneous laser field~\cite{husakou11,yavuz12, marcelo12} as produced by nano-plasmonic antennas \cite{kim08}, metallic waveguides \cite{park11}, metal~\cite{z11,s11} and dielectric~\cite{s11bis} nanoparticles or metal nanotips \cite{h06,s10,k11,k12,h12}.
The laser electric field is then no longer homogeneous in the region where the electron dynamics takes place. Very recently, studies about how HHG spectra are modified due to non-homogeneous fields, such as those present in the vicinity of a nanostructure irradiated by a short laser pulse, have been published \cite{husakou11, yavuz12, marcelo12, marcelosubmitt} leading to the so called 'plasmonic enhanced intensity' showing the increase interests such new field rises .

\begin{figure}
\resizebox{3in}{!}{\includegraphics[angle=270]{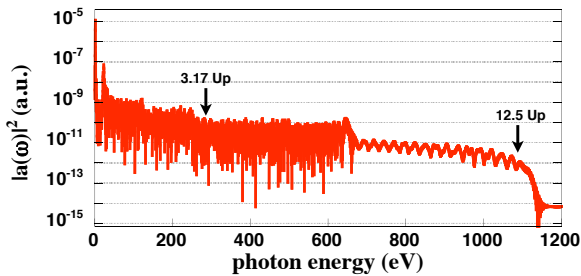}}  
\caption{3D TDSE harmonic spectrum in Helium atom generated by laser pulses described in Eq. (\ref{electric}) and ${\beta=0.002}$ for a plasmonic enhanced intensity $I=1.4\times 10^{15}$ W/cm$^2$.}
\label{fig1}
\end{figure}

In order to calculate the harmonic spectra we resolve the three dimensional Time Dependent Schr\"odinger Equation (3D-TDSE) in the length gauge while employed a double pulse and non-homogeneous driving laser field. The harmonic yield of an atom is proportional to the Fourier transform of the dipole acceleration of its active electron and it can be calculated from the time propagated electronic wave function. We have used our code which is based on an expansion in spherical harmonics, $Y_{l}^{m}$, considering only the $m = 0$ terms due to the cylindrical symmetry of the problem. The numerical technique is based on a Crank-Nicolson method implemented on a splitting of the time-evolution operator that preserves the norm of the wave function. We base our studies in He due to the fact that a majority of experiments in HHG are carried out in noble gases. Hence we have considered in our 3D-TDSE code the atomic potential reported in~\cite{lin05} to describe accurately the He atom. We remark that our scheme can be applied to any atom or molecule, once the adequate potential is chosen. The coupling between the atom and the laser pulse, linearly polarized along the $z$ axis, is modified in order to treat spatially non-homogeneous fields as follow:
\begin{eqnarray}
\label{vlaser}
V_{l}(z,t,\tau)&=&\tilde{E}(z,t,\tau)\,z
\end{eqnarray}
with 
\begin{equation}
\label{electric}
\tilde{E}(z,t,\tau)=E(t,\tau)(1+\beta z)
\end{equation}
where $E(t,\tau)$ is the laser field defined by the sum of the pulses defined in Eq. (\ref{e1}) and (\ref{e2}), the parameter $\beta$ defines the strength of the non-homogeneity and the dipole approximation is preserved because $\beta\ll1$. The harmonic spectrum then obtained in He atom for a $\beta=0.002$ is presented in Fig.~\ref{fig1}.  We can observe a considerable cut-off extension up to $12.5 U_p$ which is much greater than whilst the double pulse configuration is employed alone (that leads to a maximum of  $4.5 U_p$~\cite{perez-hernandezMD09} ). This large extension of the cut-off is therefore a signature of the combined effect of the temporal double pulse and the spatial non-homogeneous character of the laser electric field. For this particular value of laser peak intensity ($1.4\times 10^{15}$ W/cm$^{2}$) the highest photon energy is greater than 1 keV. Note that the cited intensity is actually the plasmonic enhanced intensity, not the input laser intensity. The latter could be several orders of magnitude smaller, according to the plasmonic enhancement factor (see e.g.~\cite{kim08,park11}). 

\begin{figure}
\resizebox{3.4in}{!}{\includegraphics[angle=270]{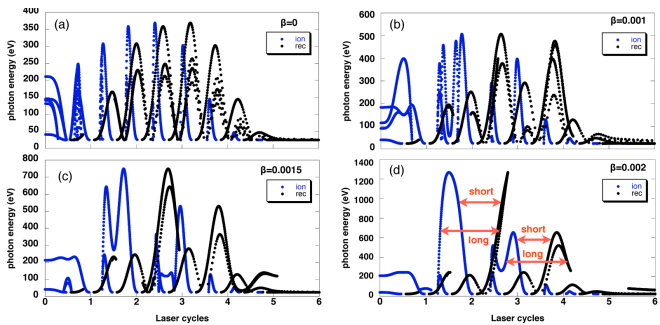}} 
\caption{Rescattering energies of electrons as a function of the ionization time (blue) and recombination time (black), for laser pulses described in Eq. (\ref{electric}) for different values of $\beta=0.0$ (a), $\beta=0.001$(b), ${\beta=0.0015}$(c) and ${\beta=0.002}$(d). In this case the plasmonic enhanced intensity is $I=1.4\times10^{15}$ W/cm$^{2}$ which corresponds to the saturation intensity for He.}
\label{fig2}
\end{figure}

In order to understand the mechanism link to this particular extension of the cut-off, we investigated the classically retrieved trajectories by solving the Newton equation of an electron in the such laser field Eq. (\ref{electric}). We calculated the energy of the returning electron as a function of the time duration of the laser pulse~\cite{marcelo12}. In Fig.~\ref{fig2} we show the classical rescattering energies (in eV) as a function of ionization time (in blue) and recombination times (in black) including the effect of the non-homogeneous field varying the $\beta$ parameter. The direct effect is then that the number of recombinations decreases with $\beta$ increases. The non-homogeneity of the laser field acts therefore as an electronics trajectories filter, with the advantage that the most energetic trajectories are selected leading to a cut-off extension greater than 1 keV. Further to this selection and extension of photon energies, an additional very interesting effect linked to the non-homogeneity of the laser field rises. Indeed for the case of $\beta=0.002$ the so-called "short" and "long"  classical trajectories~\cite{schaf93A,corku93A,zair08} recombines now almost simultaneously. In other words, the non-homogeneity of the laser field acts as a temporal lens which forces to the electrons previously ionized at different times to recombine at the same time as it is shown in the panel (d) of Fig.~\ref{fig2}. This way the attosecond chirp is significantly reduced.

\begin{figure}[here]
\resizebox{3.4in}{!}{\includegraphics[angle=270]{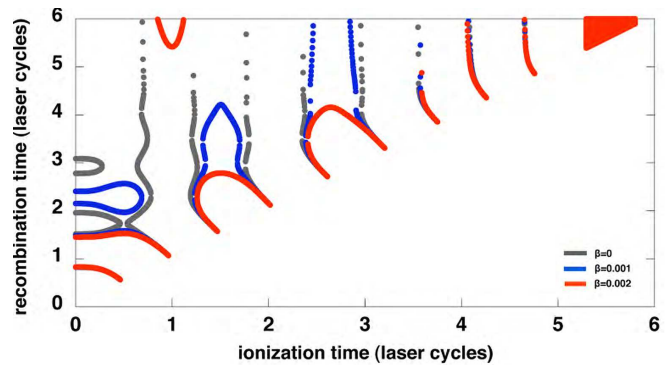}} 
\caption{Recombination times as a function of ionization times for three of the four cases exposed in Fig.~\ref{fig2} for the He atom. The case ${\beta=0}$ is plotted in grey, ${\beta=0.001}$ in blue and ${\beta=0.002}$ in red.}
\label{fig3}
\end{figure}

In addition, in Fig.~\ref{fig3} , the recollision time $t_r$ of the electron is presented as a function of the ionization time $t_i$ and for several values of $\beta$. By considering ionization times of $1.25<t_i<2.25$ the long trajectories are those with recollision times $t_r\gtrsim2.5$ optical cycles and, only for the homogeneous laser field case (grey lines in figure), these trajectories are clearly visible. On the other hand, short trajectories are characterised by $t_r\gtrsim2.5$ optical cycles and these are present for both the homogeneous and non-homogeneous cases. The results presented in here are consistent with those shown in~\cite{yavuz12,marcelo12}, but note that now the laser pulse has a more complex temporal shape. These results on the fact that the long trajectories are modified both by the non-homogeneity and the temporal double pulse configuration, namely, the homogeneous long trajectories (grey lines) with ionization times $t_i$ around 1.25 and 1.75 optical cycles merge into unique trajectories. The branch with $t_i\sim1.75$ now has ionization times more than half an optical cycle that get smaller whilst $\beta$ increases. As a result, the time spent by the electron in the continuum increases. The electric field strength at the ionization time for short trajectories being greater than for long trajectories, and considering that the ionization rate is a non-linear function of this electric field, long trajectories are then less efficient than the short ones. On the other hand, short trajectories are almost independent of $\beta$ and get noticeable different only for really high values of $\beta$. Similar behaviour can be found in Fig.~\ref{fig3} for ionization times of $2.5<t_i<3.5$.

This effect has important consequences in the harmonic spectrum plotted in Fig.~\ref{fig1} which exhibits a wide zone where no clear harmonics peaks are visible resulting in a continuum. As demonstrated in~\cite {sansone_s_2006}, for continuum obtained around 90 eV using a 800 nm laser field using single few-cycle pulse and homogeneous field, continuum generation in the harmonics spectra is a necessary condition for the generation of an isolated attosecond pulse. So far no clear method has been proposed to allow the generation of isolated attosecond pulse in the keV region. Therefore generating harmonics using the combine effect of temporal and spatial synthesis is a completely new route towards the first generation of isolated attosecond pulses beyond the Carbon K-edge.

\begin{figure} [here]
\resizebox{3.4in}{!}{\includegraphics[angle=270]{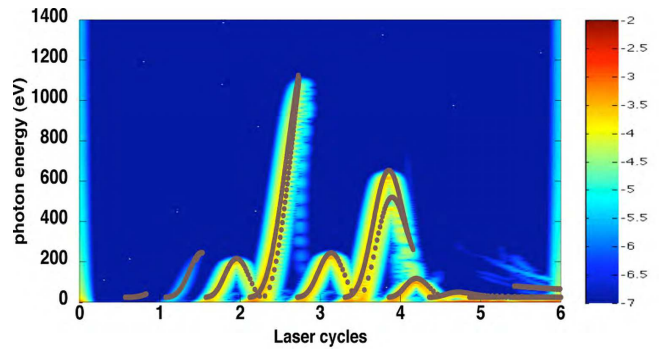}} 
\caption{Time-frequency analysis obtained from the full integration of the 3D TDSE and superimposed (in brown) classical rescattering energies for the case of ${\beta=0.002}$ and a plasmonic enhanced intensity $I=1.4 \times 10^{15}$ W/cm$^2$ corresponding to the case of Fig.~\ref{fig2}.}
\label{fig4}
\end{figure}

As a final test to confirm the underlying physics highlighted by the classical trajectories analysis, we retrieved the time-frequency distribution of the calculated dipole (from the 3D-TDSE) corresponding to the case of a non-homogeneous laser field using a wavelet analysis. The result is presented in Fig.~\ref{fig4} where we have superimposed the classical rescattering energies (in brown) to show how the two calculations match. Analogously, in Fig.~\ref{fig4} we show the time-frequency analysis in the case of $\beta=0.002$ that corresponds to the spectra presented in Fig.~\ref{fig1}. The consistency of the classical calculations with the full quantum approach is clear and confirms that the mechanism of the generation of this $12.5 U_p$ cut-off extension exhibiting a nice continuum is the consequence of trajectory selection and temporal lens effect on the recollision time whilst employing the combination of temporal and spatial synthesised laser field.

In conclusion, using a laser fields with a temporal superposition of two identical few-cycle pulses delayed in time together with a spatial non-homogeneity, we demonstrate that temporally and spatially synthesised laser field can lead to a new route for hight harmonic generation, attosecond pulse production and in a further extend to strong field physics.  We demonstrate that the main effect of this synthesised laser field on the harmonics results in a considerable extension of the cut-off energy up to $12.5 U_p$  which present a nice continuum generation compatible with the generation of isolated attosecond pulse beyond the Carbon K-egde. This effect is understood by analysing the trajectories involved in the process using a classical and a full quantum approaches. Both analysis converge to the same conclusion: trajectories are highly selected whilst using a laser field that consists of a combination of the double pulse temporal synthesis and the spatially non-homogeneity. And a 'temporal lens effect' shrink the recombination time window, resulting in a nice continuum generation in the extended harmonics spectral region. This new approach provides a unique route to the generation of coherent attosecond light source beyond the Carbon K-edge directly from input 800 nm laser system and therefore is of interest of strong field applications time-resolved  X-ray spectroscopy of molecular system.    

\begin{acknowledgments}
 J. A. P.-H. and L. R. acknowledge support from Spanish MINECO through the 
Consolider Program SAUUL (CSD2007-00013) and research project FIS2009-09522, from 
Junta de Castilla y Le\'on through the Program for Groups of Excellence (GR27) and from the 
ERC Seventh Framework Programme (LASERLAB-EUROPE, grant agreement n 228334);
M. F. C. and M. L. acknowledge the financial support of the MINCIN projects (FIS2008-00784 TOQATA and Consolider Ingenio 2010 QOIT); M. L. acknowledges the financial support of the ERC Advanced Grant QUAGATUA, Alexander von Humboldt Foundation and Hamburg Theory Prize.
A. Z, acknowledge the UK EPSRC support through project Grant No. EP/J002348/1.

\end{acknowledgments}



\begin{thebibliography}{}

\bibitem{rehr00}J.J. Rehr \textit{et al.}, Rev. Mod. Phys. {\bf 72} 621 (2000).
\bibitem{bressler04} C. Bressler and M. Chergui, Chem. Rev. {\bf 104} 1781 (2004).
\bibitem{lewenstein94A} M. Lewenstein et al.,  Phys. Rev. A {\bf 49} 2117 (1994) 
\bibitem{pop12} T. Popmintchev et al., Science {\bf 336}, 1297-1291 (2012).
\bibitem{tate_07A} J. Tate \textit{et al.}, Phys. Rev. Lett. {\bf 98}, 013901 (2007).
\bibitem{frolov_08} M. V. Frolov, N. L. Manakov, and A. F. Starace, Phys. Rev. Lett. {\bf100}, 173001-1-4 (2008).
\bibitem{perez-hernandez09A} J. A. P\'erez-Hern\'andez, L. Roso, L. Plaja, Opt. Express {\bf 17}, 9891 (2009)
\bibitem{zeng07} Z. Zeng \textit{et al.}, Phys. Rev. Lett. {\bf 98}, 203901 (2007).
\bibitem{Takahashi08} E. J. Takahashi \textit{et al.},  Phys. Rev. Lett. {\bf 101}, 253901 (2008)
\bibitem{Winterfeldt08} C. Winterfeldt, C. Spielmann, and G. Gerber, Rev. Mod. Phys. {\bf 80}, 117 (2008).
\bibitem{chipperfield09} L. E. Chipperfield \textit{et al.}, Phys. Rev. Lett. {\bf 102}, 063003 (2009)
\bibitem{chipperfield10} L. E. Chipperfield, J. W. Tisch, and J. P. Marangos, J. Mod. Opt.  {\bf 57}, 992 (2010).
\bibitem{siegel10} T. Siegel \textit{et al.}, Opt. Express {\bf 18}, 6853 (2010)
\bibitem{brugera10} L. Brugnera \textit{et al.}, Opt. Lett. {\bf 35}, 3994 (2010).
\bibitem{Orlando2010} G. Orlando; P. P. Corso, and E. Fiordilino, J. Mod. Opt.  {\bf 57}, 2069 (2010). 
\bibitem{brugera11} L. Brugnera \textit{et al.}, Phys. Rev. Lett.  {\bf 107}, 153902 (2011).
\bibitem{ganeev12}	R. A. Ganeev \textit{et al.} Phys. Rev. A  {\bf 85}, 023832 (2012).
\bibitem{bandrauk12} K. Yuan and A. D Bandrauk, J. Phys. B {\bf 45}, 074001 (2012)		
\bibitem{carrera09} J.J. Carrera and X. M. Tong, Phys. Rev. A {\bf 74}, 023404 (2006)
\bibitem{perez-hernandezMD09}  J. A. P\'erez-Hern\'andez et al., J. Phys. B  {\bf 42}, 134004 (2009) 
\bibitem{sansone_s_2006} G. Sansone \textit{et al.}, Science \textbf{314}, 443 (2006).
\bibitem{eckle_np_2008} P. Eckle \textit{et al.}, Nat. Phys. \textbf{4} 565-570 (2008).
\bibitem{husakou11} A. Husakou, S.-J. Im, and J. Herrmann, Phys. Rev. A {\bf 83}, 043839 (2011) 
\bibitem{yavuz12} I. Yavuz, Phys. Rev. A {\bf 85}, 013416 (2012).
\bibitem{marcelo12} M. F. Ciappina \textit{et al.}, Phys. Rev. A {\bf 85}, 033828 (2012).
\bibitem{protopapas97} M. Protopapas, C. H. Keitel, and P. L. Knight, Rep. Prog. Phys. {\bf 60}, 389 (1997)
\bibitem{brabec00} T. Brabec and F. Krausz, Rev. Mod. Phys. {\bf 72}, 545 (2000)
\bibitem{strelkov09} V. Strelkov, E. Mevel and E. Constant, Eur. Phys. J. Special Topics {\bf 175}, 15-20 (2009).
\bibitem{kim08}S. Kim, J. Jin \textit{et al.}, Nature (London) {\bf 453}, 757 (2008).
\bibitem{park11}I.-Y. Park \textit{et al.}, Nat. Phot. {\bf 5}, 677 (2011).
\bibitem{z11} S. Zherebtsov \textit{et al.}, Nat. Phys. {\bf 7}, 656 (2011).
\bibitem{s11}  F. S\"{u}{\ss}mann and M. F. Kling, Proc. of SPIE {\bf 8096}, 80961C (2011).
\bibitem{s11bis} F. S\"{u}{\ss}mann and M. F. Kling, Phys. Rev. B {\bf 84}, 121406(R) (2011).
\bibitem{h06}  P. Hommelhoff \textit{et al.}, Phys. Rev. Lett. {\bf 96}, 077401 (2006).
\bibitem{s10} M. Schenk, M. Krueger, and P. Hommelhoff, Phys. Rev. Lett. {\bf 105}, 257601 (2010).
\bibitem{k11}  M. Krueger, M. Schenk, and P. Hommelhoff, Nature {\bf 475}, 78 (2011).
\bibitem{k12} M. Krueger \textit{et al.}, J. Phys. B {\bf 45}, 074006 (2012).
\bibitem {h12} G. Herink \textit{et al.}, Nature {\bf 483}, 190 (2012).
\bibitem{marcelosubmitt} M. F. Ciappina \textit{et al.}, Phys. Rev. A., submitted (2012).
\bibitem{lin05} X. M. Tong and C. D. Lin, J. Phys. B: At. Mol. Opt. Phys. {\bf 38}, 2593 (2005).
\bibitem{schaf93A} K. Schafer \textit{et al.}, Phys. Rev. Lett. {\bf 70}, 1599 (1993).
\bibitem{corku93A} P. B. Corkum,  Phys. Rev. Lett. {\bf 71}, 1994 (1993).
\bibitem{zair08} A. Za\"ir \textit{et al.}, Phys. Rev. Lett {\bf 100}, 143902 (2008).
%
%
%
%
%
%
%
%
%
%
%
%
%
%
%
%
%
%
%
%







%
%
%

%
%
%
%
%

%
%
%
 
\end{thebibliography}

\end{document}